\documentclass[journal]{IEEEtran}
\ifCLASSINFOpdf
  \usepackage[pdftex]{graphicx}
  \graphicspath{{../pdf/}{../jpeg/}}
  \DeclareGraphicsExtensions{.pdf,.jpeg,.png}
\else
  \usepackage[dvips]{graphicx}
  \graphicspath{{../eps/}}
  \DeclareGraphicsExtensions{.eps}
\fi
\usepackage{epsfig}
\usepackage{booktabs}
\usepackage{epstopdf}
\usepackage{multirow}
\usepackage{amssymb}
\usepackage{mathtools}
\usepackage{subfigure}
\usepackage{adjustbox}
\usepackage{bbm}

\usepackage{algorithmic}

\usepackage{array}
 \usepackage{mathrsfs}
\usepackage{svg}
\usepackage{amssymb}
\usepackage{amsmath}
\usepackage{cite}
\usepackage{url}
\usepackage{xcolor}
\usepackage{cite,graphicx,amsmath,amssymb}
\usepackage{subfigure}
\usepackage{makecell}
\usepackage{fancyhdr}
\usepackage{mdwmath}
\usepackage{mdwtab}
\usepackage{amsthm}
\usepackage{suffix}
\usepackage{mathtools}
\allowdisplaybreaks[4]
\usepackage{xpatch}
\usepackage{graphicx}
\usepackage{eucal}
\usepackage{enumitem} 
\newlist{steps}{enumerate}{1}
\setlist[steps, 1]{label = Step \arabic*:}

\xpatchbibmacro{name:andothers}{%
  \bibstring{andothers}%
}{%
  \bibstring[\emph]{andothers}%
}{}{}

\DeclarePairedDelimiterX\MeijerM[3]{\lparen}{\rparen}%
{\begin{smallmatrix}#1 \\ #2\end{smallmatrix}\delimsize\vert\,#3}

\newcommand\MeijerG[8][]{%
  G^{\,#2,#3}_{#4,#5}\MeijerM[#1]{#6}{#7}{#8}}

\WithSuffix\newcommand\MeijerG*[7]{%
  G^{\,#1,#2}_{#3,#4}\MeijerM*{#5}{#6}{#7}}

\newtheorem{theorem}{Theorem}

\newtheorem{lemma}{Lemma}

\newtheorem{corollary}{Corollary}

\begin{document}
\title{Applying Deep-Learning-Based Computer Vision to Wireless Communications: Methodologies, Opportunities, and Challenges 
\thanks{Manuscript received on August 3, 2020; revised on November 1, 2020 and November 30; accepted on December 1, 2020 by IEEE Open Journal of the Communications Society. The associate editor coordinating the review of this paper and approving it for publication was Dr. Gayan Aruma Baduge. (Corresponding author: Gaofeng Pan.)}
\thanks{Authors are with Computer, Electrical and Mathematical Sciences and Engineering Division, King Abdullah University of Science and Technology (KAUST), Thuwal 23955-6900, Saudi Arabia.}}

\author{Yu Tian, Gaofeng Pan,~\IEEEmembership{Senior Member,~IEEE}, and Mohamed-Slim Alouini,~\IEEEmembership{Fellow,~IEEE}}

\maketitle
\begin{abstract}
 
Deep learning (DL) has seen great success in the computer vision (CV) field, and related techniques have been used in security, healthcare, remote sensing, and many other areas. As a parallel development, visual data has become universal in daily life, easily generated by ubiquitous low-cost cameras. Therefore, exploring DL-based CV may yield useful information about objects, such as their number, locations, distribution, motion, etc. Intuitively, DL-based CV can also facilitate and improve the designs of wireless communications, especially in dynamic network scenarios. However, so far, such work is rare in the literature. The primary purpose of this article, then, is to introduce ideas about applying DL-based CV in wireless communications to bring some novel degrees of freedom to both theoretical research and engineering applications. To illustrate how DL-based CV can be applied in wireless communications, an example of using a DL-based CV with a millimeter-wave (mmWave) system is given to realize optimal mmWave multiple-input and multiple-output (MIMO) beamforming in mobile scenarios. In this example, we propose a framework to predict future beam indices from previously observed beam indices and images of street views using ResNet, 3-dimensional ResNext, and a long short-term memory network. The experimental results show that our frameworks achieve much higher accuracy than the baseline method, and that visual data can significantly improve the performance of the MIMO beamforming system. Finally, we discuss the opportunities and challenges of applying DL-based CV in wireless communications.
\end{abstract}


\section{Introduction}

Recently, deep learning (DL) has seen great success in the computer vision (CV) field. DL networks comprise networks such as deep neural networks, deep belief networks, recurrent neural networks (RNNs), and convolutional neural networks (CNNs). Many DL networks with various structures have emerged with the availability of large image and video datasets and high-speed graphic processing units (GPUs) \cite{goodfellow2016deep}. DL networks can achieve success in CV because they discover and integrate low-/middle-/high-level features in images and leverage them to accomplish specific tasks \cite{he2016deep}. DL can easily fulfill CV applications with remarkably high performance, such as semantic segmentation, image classification, and object detection/recognition \cite{goodfellow2016deep}. DL-based CV has therefore been widely utilized in public security, healthcare, and remote sensing, as such fields generate much visual data \cite{bharti2020recent}. {However, DL-based CV is rarely seen in the design and optimization of wireless communication systems in which the researchers mainly focus on the transmission quality of the information bits/packets, e.g., transmission rate, bit/packet error, traffic/user fairness, etc. via purely exploiting the information on the transmission behaviors of radio frequency signals (e.g., the power, direction, phase, transmission duration, etc.), rather than making use of the geometry information of the surrounding space.} Thus, such presented design and optimization of wireless communications cannot achieve the optimal performance with no doubts.

Nowadays, high-definition cameras are installed almost everywhere because of their low cost and small size. In some public areas, cameras have long existed for monitoring purposes. Therefore, visual data can easily be obtained in wireless communication systems in real life \cite{xu20193d}. As useful information about \textit{static system topology} (including terminals' numbers, positions, distances among themselves, etc.) and \textit{dynamic system information} (including moving speed, direction, and changes in the number of the terminals) can be recognized, estimated, and extracted from these multi-medium data via DL-based CV techniques, new potential benefits can be exploited for wireless communications to aid system design/optimization, such as resource scheduling and allocations, algorithm design, and more.

\begin{figure*}[!htbp]
    \setlength{\abovecaptionskip}{0pt}
    \setlength{\belowcaptionskip}{10pt}
     \centering
    \includegraphics[width=5.5 in]{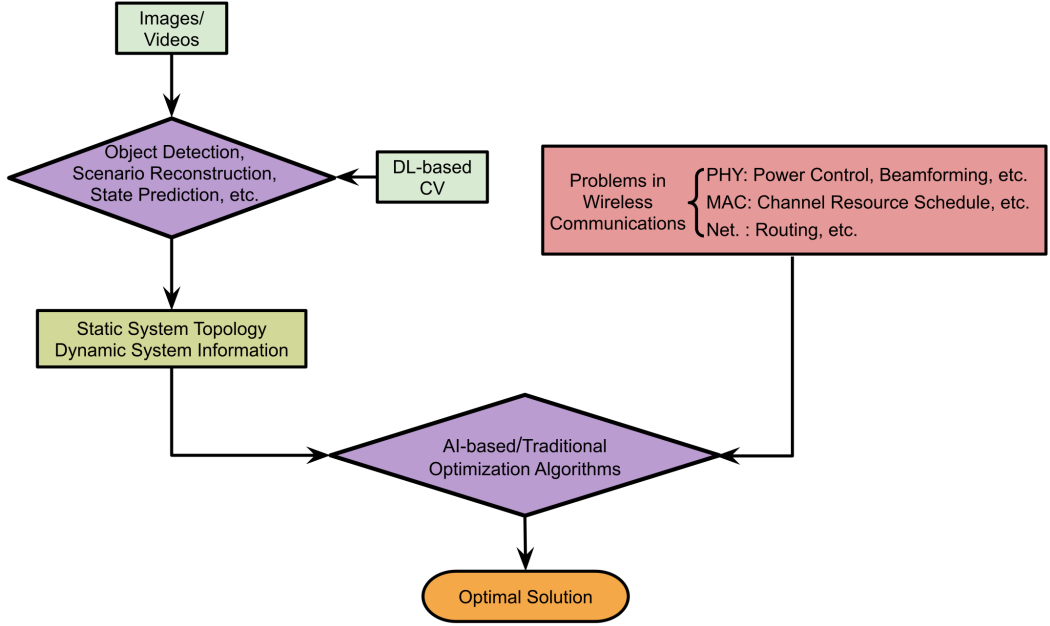}
    \caption{Framework of applying DL-based CV to wireless communications (PHY: Physical layer; MAC: MAC layer; Net.: Network layer; AI: artificial intelligence)}
    \centering
    \label{fig:flowchart}\vspace{-2mm}
\end{figure*}

Fig. \ref{fig:flowchart} presents the framework of applying DL-based CV to wireless communications, the core idea of which is to explore the useful information obtained/forecasted by DL-based CV techniques to facilitate the design of wireless communications via DL-based/traditional optimization methods. In the following, we introduce some applications of DL-based CV in wireless systems in three aspects: the physical layer, medium access control (MAC) layer, and network layer. 

{1) In the physical layer of wireless communication systems, traditional methods usually first estimate the channel state by sending pilot signals from the transmitter to receivers \cite{zhou2004adaptive}. Then according to the achieved channel state information (CSI), specific modulation, source encoding, channel encoding, and power control strategies can be selected to realize the optimal utilization of system resources (e.g., bandwidth and energy budgets). However, the CSI only contains amplitudes and phases information of the channel fading rather than the locations, number, and environmental information of the users which can be easily obtained from visual data by object detection and segmentation techniques in CV, leading to the fact that the real optimal system performance cannot be realized. However, with the aid of the more comprehensive users' information, dynamic modulation, encoding, and power control can be easily and optimally formulated and implemented. For example, in multiple-input and multiple-output (MIMO) beamforming communication systems, the direction and power of beams can be scheduled using the knowledge of users' locations and blocking cases in the visual data, which cannot be obtained via traditional methods. }

{2) In the MAC layer, like in cellular wireless networks, receiver-to-transmitter feedback information and cell-to-cell CSI is very important information to be utilized to allocate resources and to guarantee the quality of service in the traditional methods \cite{gesbert2007adaptation}. Thus, a long time delay may always exist when analyzing the feedback and CSI in crowded scenarios in which a huge number of users are served by the network.} By jointly using these information and the density or distribution of users obtained from the visual data in the serving area of the BS, channel resources (including frequency bands, time slots, etc.) can be efficiently reserved and allocated to achieve optimal overall performance. For example, smart homes have various kinds of terminals such as smartphones, televisions, laptops, and other intelligent home appliances. As such, channel resources can be dynamically scheduled by considering the information obtained from the visual data, such as the number and locations of the users. {Unlike traditional handover algorithms that adopt the measured fluctuation of received signal power to estimate the distance between the terminal and BS, the moving information including velocity and its variations can be fully estimated from visual data to accurately facilitate channel resource allocation in the handover process.} This will be quite useful in fifth-generation wireless networks due to the shrinking sizes of the serving zones.

{3) For the network layer, taking multi-hop transmission scenarios as an example, traditional routing algorithms are mostly running based on the length of the routing path estimated by the pilot and feedback signals which cannot reflect the timely location changes in mobile scenarios \cite{jung2017topology}. By exploiting system topology information from the visual data, novel routing algorithms can be designed to efficiently improve transmission performance, such as the end-to-end delivery delay, packet loss rate, jam rate, and system throughput.} For another instance, wireless sensor networks have numerous sensors that can be deployed in target areas to monitor, gather, and transmit information about their surrounding environments. Then, the system topology information from visual data can be used to design multi-hop transmissions, which are required due to the inherent resource limitations and hardware constraints of the sensors.

In general, traditional algorithms adopted in wireless communication systems depend on traditional channel/network state estimation methods to grab the CSI and network state information which unavoidably suffer time delay and/or feed errors, resulting in low efficiency or even wrong decisions. Especially, it is hard or impossible to get accurate CSI or network state information in high dynamic network scenarios through traditional methods. Thanks to the inherent merits of DL-based CV techniques, the static and dynamic system information can be accurately and efficiently extracted from visual data, bringing vital benefits to the design and optimization of wireless communication systems. 

In this context, this article introduces the methodologies, opportunities, and challenges of applying DL-based CV in wireless communications as an essential reference/guide for theoretical research and engineering applications. 

The rest of this article is organized as follows. Section II overviews related work from two perspectives: 
datasets and applications. Section III presents an example of applying a DL-based CV to mmWave MIMO beamforming and elaborates on the problem definition, framework architecture, pipeline, the results of the example, and practical application. Section IV introduces and discusses some challenges and open problems of applying DL-based CV to wireless communications. Finally, Section V concludes the article. 

\section{An Overview of Related Work}
Applying DL-based CV to wireless communications has two essential dimensions: datasets and applications. In the following, we give a brief overview of recent work in these two aspects.

\paragraph{Datasets}Building datasets is an essential step as DL is data-hungry. In \cite{alrabeiah2019viwi}, the authors proposed a parametric, systematic, scalable dataset framework called Vision-Wireless (ViWi). They utilized this framework to build the first-version dataset containing four scenarios with different camera distributions (co-located and distributed) and views (blocked and direct). These scenarios were based on a millimeter wave (mmWave) MIMO wireless communication system. Each scenario contained a set of images captured by the cameras and raw wireless data (signal departure/arrival angles, path gains, and channel impulse responses). Using the provided MATLAB script, they could view the user's location and channel information in each image from the raw wireless data. Later, the same authors built the second-version dataset called ViWi Vision-Aided Millimeter-Wave Beam Tracking (ViWi-BT) \cite{alrabeiah2020viwi} and posted it for the machine learning competition at the IEEE International Conference on Communications (ICC) 2020. This dataset contains images captured by the co-located cameras and mmWave MIMO beam indices under a predefined codebook. Section \ref{dataset} covers the details of this dataset. {The authors of \cite{Klautau20185g} introduced another dataset called Raymobtime which contains ray-tracing, LIDAR, matrix channel, GPS, and image data in mmWave MIMO  vehicle-to-infrastructure wireless communication systems. Notably, ray-tracing data provides path parameters such as received power, time of arrival, angle of departure, angle of arrival, line of a sight ray status, and ray phase while GPS user info data has line-of-sight (LOS) status, channel valid or not information, number of the TX in the vehicle, and the 3D coordinates.} Beam selection and channel estimation challenges are held based on this dataset in International Telecommunication Union (ITU) Artificial Intelligence/Machine Learning in 5G Challenge. {As both ViWi and Raymobtime datasets are versatile, except the challenges in ICC and ITU, many other interesting applications (e.g., blockage prediction, power prediction, angle estimation, LOS decision, etc.) can be explored according to their abundant data.} In \cite{ayvacsik2019veni}, a dataset consisting of depth image frames from recorded videos was built and can be applied in channel estimation tasks.

\paragraph{Applications} There are plenty of interesting applications of DL-based CV techniques designed to tackle problems in wireless communications. A framework to implement beam selection in mmWave communication systems by leveraging environmental information was presented by \cite{xu20193d}. The authors used the images with different perspectives captured by one camera to construct a three-dimensional (3D) scene and generate corresponding point cloud data. They built a model based on 3D CNN to learn the wireless channel from the point cloud data and predict the optimal beam. Based on the first-version ViWi dataset, \cite{alrabeiah2019millimeter} proposed a modified ResNet18 model to conduct beam and blockage prediction from the images and channel information. Based on the second-version ViWi-BT dataset, the authors of \cite{alrabeiah2020viwi} provided a baseline method based on Gated Recurrent Units (GRUs) without the images, only the beam indices. They believe they can achieve better performance if they leverage both kinds of data. 
Based on the Raymobtime dataset, CNN and deep reinforcement learning (DRL) were utilized to select proper pair of beams for vehicles with images generated from GPS locations data in vehicle-to-infrastructure scenarios in \cite{Klautau20185g}. The authors also compared DL-based methods with other traditional machine learning methods such as SVM, AdaBoost, decision tree, and random forest. The results showed that DL-based method has the best performance. In \cite{Klautau2019lidar}, two CNNs were proposed to conduct line-of-sight decision and beam selection by using LIDAR point cloud data in the Raymobtime dataset. {In \cite{charan2020vision}, authors proposed a neural network containing CNNs and an RNN-based recurrent prediction network to predict the dynamic link blockages using red, green, blue (RGB) images and beamforming vectors provided by the extended ViWi-BT dataset. In \cite{ayvacsik2019veni}, authors developed a CNN-based framework called VVD to estimate the wireless communication channels only from only depth images in mmWave systems. In \cite{nishio2019proactive}, a framework consisting of CNN and convolutional LSTM (convLSTM) network was presented to proactively predict the received power through depth images in mmWave networks and exhibited the highest accuracy compared with the random forest algorithm and a CNN-based method. In \cite{koda2020handover}, a proactive handover management framework was proposed to make handover decisions by using camera images and DRL. In \cite{koda2020commu}, a multimodal split learning method based on convLSTM networks was presented to predict mmWave received power through camera images and radio frequency signals while considering communication efficiency and privacy protection. All aforementioned papers are summarized in Table \ref{paperreviewtable} for comparison purposes.}
\begin{table*}[!ht]
 \caption{{Paper review}}
 \label{paperreviewtable}
\resizebox{\textwidth}{!}{\begin{tabular}{c|c|c|c|m{.1\textwidth}<{\centering}}
\hline \hline
Paper & Task           & Network & Data type & Dataset public available?   \\\hline
\cite{xu20193d}& Beam selection & 3D CNN  &   Point clouds generated from camera images& No \\\hline
\cite{alrabeiah2019millimeter}&Beam and blockage prediction &Modified ResNet18& Camera images&Yes\\\hline
\cite{alrabeiah2020viwi}&Beam prediction&GRU& Previous beams&Yes\\\hline
\cite{Klautau20185g}& Beam selection&CNN and DRL& Images generated from vehicle locations&Yes\\\hline
\cite{Klautau2019lidar}&Beam selection&  CNN& LIDAR point clouds& Yes\\\hline
\cite{charan2020vision}&Blockage prediction&CNN and RNN         &Camera images and previous beams&Yes\\\hline
\cite{ayvacsik2019veni}& Channel estimation& CNN&Depth images&Yes\\\hline
\cite{nishio2019proactive}&Received power prediction&CNN and convLSTM network&Depth images&No\\\hline 
\cite{koda2020handover}&Handover decision &DRL&Camera images&No\\\hline
\cite{koda2020commu}&Received power prediction&Multimodal split learning and convLSTM networks&Camera images and radio frequency signals&No\\\hline\hline
\end{tabular}}
\end{table*}

\begin{figure}[!htbp]
    \setlength{\abovecaptionskip}{0pt}
    \setlength{\belowcaptionskip}{10pt}
    \includegraphics[width=3.4 in]{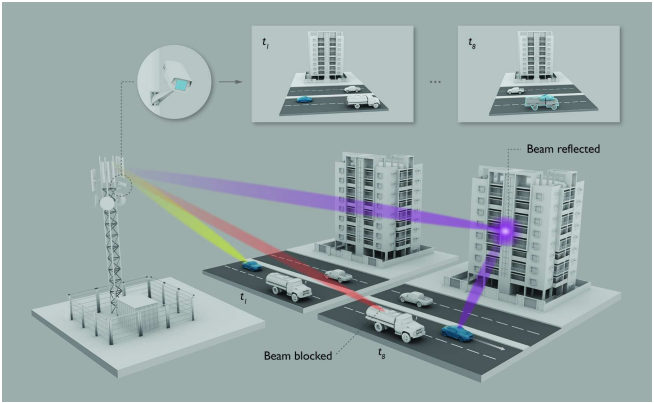}
    \centering
    \caption{Scenario of applying DL-based CV to mmWave MIMO beamforming}
    \centering
    \label{fig:system}\vspace{-0mm}
\end{figure}

\section{An Example of Applying DL-based CV to Beamforming}
\begin{figure*}[!htbp]
    \setlength{\abovecaptionskip}{0pt}
    \setlength{\belowcaptionskip}{10pt}
    \includegraphics[width=4.8 in]{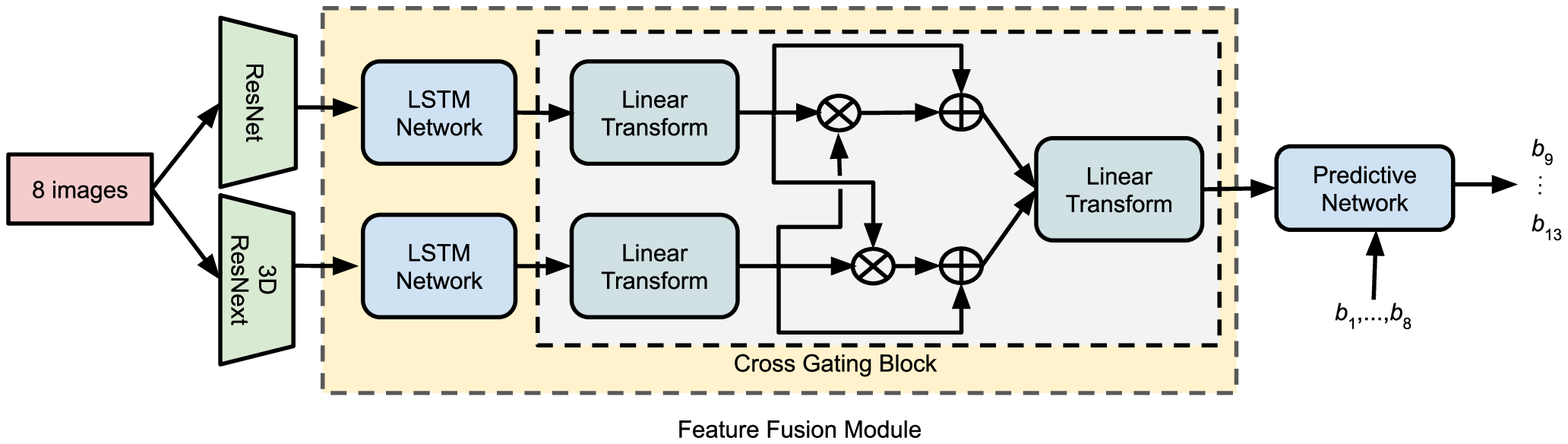}
    \centering
    \caption{Architecture of our proposed framework}
    \label{fig:framework}\vspace{-2mm}
\end{figure*}
\subsection{Problem Definition}

MmWave communication is a promising technique in the fifth-generation communication system, thanks to its broad available bandwidth and ultra-high data-transmitting rate \cite{alrabeiah2019millimeter,alrabeiah2019viwi,alrabeiah2020viwi}. MIMO and beamforming are widely used in mmWave communication systems and should be implemented in a large antenna array to achieve the required high power gain and direction. The classic beamforming and beam tracking algorithms suffer a common disadvantage: complexity increases dramatically with the number of antennas, resulting in substantial computational overhead. DL-based CV is a promising candidate to address this overhead issue.

In this section, we give an example of applying the DL-based CV to mmWave MIMO beamforming. As defined in \cite{alrabeiah2020viwi}, the considered scenario contains 2 BSs located on the opposite sides of a street with a distance of 60 meters. As shown in Fig. \ref{fig:system}, each BS forms a MIMO beam to serve a target user moving along the street. Therefore, the beam direction must be dynamically adjusted to catch the target mobile user. The target user may be blocked at some moments, such as $t_8$ in Fig. \ref{fig:system}, and then the beam cannot directly reach the target user, while proper reflection from other objects, such as buildings and vehicles, must be designed. Meanwhile, three cameras installed at the BS capture RGB images of the whole street view to assist the beamforming process. {The problem here is how to utilize the eight pairs of previously-observed consecutive beams and corresponding images to predict the future one, three, and five beams through DL model. Notably, these beams are represented as beam indices under the same predefined codebook.}

{A sequence containing the eight pairs of previously-observed images and corresponding beam indices for the $u$th user at the time instance $t$ is given as
\begin{align}
    S_{u}[t]=\{(X_{u}[i],b_u[i])\}_{i=t-7}^{t},
\end{align}
where $X_u[i]$ is the RGB image taken at the $i$th time instance and $b_u[i]$ is the corresponding beam index.}

{Let $f_{\Theta}(S_u[t])$ be a prediction function of a DL model and $\Hat{b}_u[t+n]$ ($n=1,...,5$) be the predicted beam index at the time instance $t+n$. $f_{\Theta}(S_u[t])$ takes in the sequence $S_{u}[t]$ and outputs a predicted sequence $\{\Hat{b}_u[t+n]\}_{n=1}^{5}$. $\Theta$ is a set of parameters of the DL model which is obtained by training the model with the training set. The training set consists of labelled sequences, i.e., $\mathcal{D}=\{(S_u[t],\{g_u[t+n]\}_{n=1}^{5})\}_{u=1}^{U}$ where each pair consists of an observed sequence and five groundtruth future beam indices.}

{Equivalent to the defined problem, our goal is to get the prediction function which can maximize the joint success probability of all data samples in $\mathcal{D}$. The object function is expressed as 
\begin{alignat}{2}
\max_{f_{\Theta}(S_u[t])} \quad  \prod_{u=1}^{U}\prod_{n=1}^{5} {\bf{Pr}}\left\{\left.\Hat{b}_u[t+n]=g_u[t+n]\right|S_u[t]\right\},
\end{alignat}
where each success probability only relies on its observed sequence $S_u[t]$.}

\subsection{ Framework Architecture and Methods}\label{sectionarchitecture}

We propose a DL network framework shown in Fig. \ref{fig:framework} composed of ResNet\cite{he2016deep}, 3D ResNext\cite{hara2018can}, a feature-fusion module (FFM)\cite{wang2019controllable}, and a predictive network.
\subsubsection{ResNet, ResNext and 3D ResNext}
\begin{figure}[!htbp]
    \setlength{\abovecaptionskip}{0pt}
    \setlength{\belowcaptionskip}{10pt}
    \includegraphics[width=2.8 in]{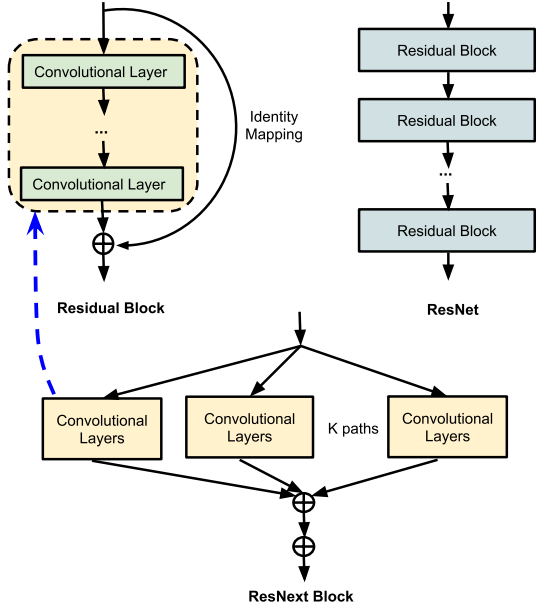}
    \centering
    \caption{Structure of residual block, ResNet, and ResNext block}
    \centering
    \label{fig:resnet}\vspace{-0mm}
\end{figure}

ResNet consists of several residual blocks, as presented in Fig. \ref{fig:resnet}. Each block contains two or more convolutional layers and superimposes its input to its output through identity mapping. It can efficiently address the vanishing gradient issue caused by the rising number of convolutional layers. If a specific number of such blocks are concatenated, as depicted in Fig. \ref{fig:resnet}, ResNet is available to achieve as many as 152 layers.

Fig. \ref{fig:resnet} also presents the structure of the ResNext block\cite{xie2017aggregated}, an improved version of the residual block, that adds a `next' dimension, also called `Cardinality'. It sums the outputs of $K$ parallel convolutional layer paths that share the same topology and inherits the residual structure of the combination. As $K$ diversities are achieved by $K$ paths, this block can focus on more than one specific feature representations of the images. 

In 3D ResNext, a similar structure can be observed but with 3D convolutional layers instead of two-dimensional (2D) ones. The 3D convolutional layer is designed to capture spatiotemporal 3D features from raw videos. 

ResNet and 3D ResNext have been widely used as feature extractors for their powerful feature-representation abilities. If they are used in a DL network directly, however, the training time will become extremely long, and many computational resources will be occupied due to the large number of layers. Therefore, researchers commonly apply a pre-trained ResNet on the ImageNet dataset to extract visual features from images and a 3D ResNext on the Kinetics dataset to extract spatiotemporal features from videos \cite{park2019adversarial}. These features are then fed to the DL network as inputs. 

\subsubsection{ Long Short-Term Memory (LSTM) Network}\label{lstms}
\begin{figure*}[!htbp]
    \setlength{\abovecaptionskip}{0pt}
    \setlength{\belowcaptionskip}{10pt}
    \includegraphics[width=6.3 in]{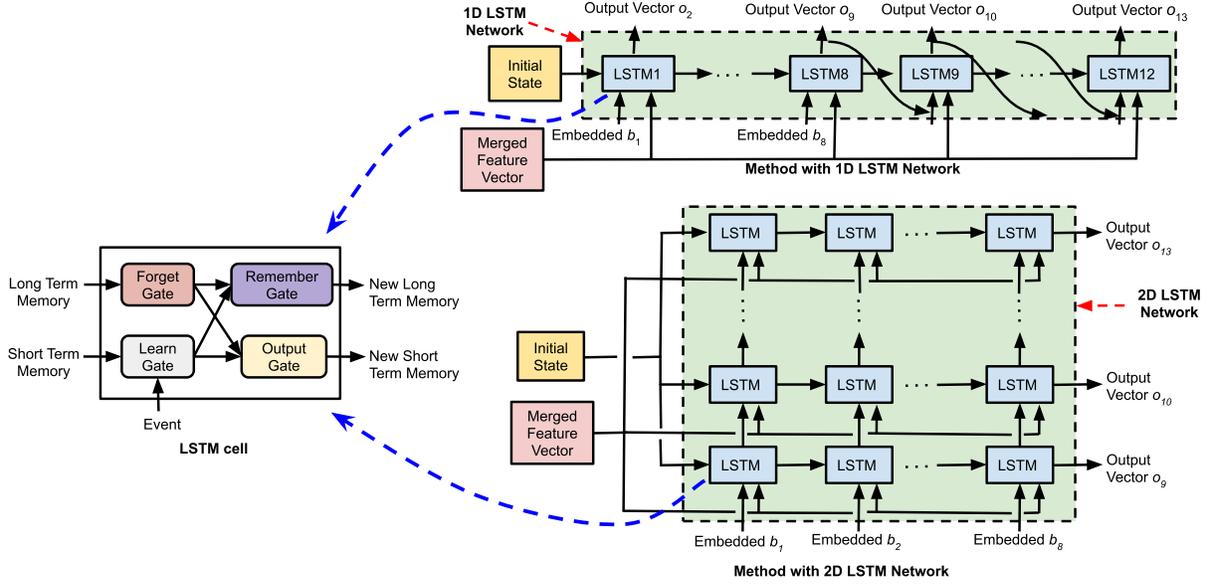}
    \centering
    \caption{ Structure of LSTM cell, method with 1D LSTM network, and method with 2D LSTM network }

    \label{fig:lstms}\vspace{-2mm}
\end{figure*}
The LSTM\cite{hochreiter1997long} network is designed for the tasks that contain time-series data, such as prediction, speech recognition, text generation, etc. Hence, it is a suitable candidate for our predictive network. The network comprises several LSTM cells, as depicted in Fig. \ref{fig:lstms}. Event (current state), previous long-term memory (cell state), and previous short-term memory (hidden state) are the inputs of an LSTM cell, in which learn, forget, remember, and output gates are employed to explore the information from the inputs. The LSTM cell outputs new long-term memory and short-term memory in which the latter is also regarded as a prediction. 

When an LSTM cell is recursively utilized in a 1D array form, a 1D LSTM network is obtained, as presented in Fig. \ref{fig:lstms}. At each moment, the cell and hidden states of the previous moment are used to generates the outputs of the current moment. 

As shown in Fig. \ref{fig:lstms}, a 2D LSTM network can be realized when the LSTM cell is recursively in a 2D mesh form \cite{bahar2018towards}. Each LSTM cell utilizes the hidden and cell states from the two neighboring cells in the left and below positions in the mesh, and its states are delivered to its neighboring cells in the right and top positions. Obviously, the number of predictions is equal to the number of rows.

\subsubsection{Feature-Fusion Module (FFM)}

 Fig. \ref{fig:framework} shows the structure of the FFM which comprises two LSTM networks and a cross-gating block. The features from ResNet and 3D ResNext are aggregated by the LSTM networks and then high-level features are obtained. The cross-gating block can make full use of the related semantic information between these two kinds of features by multiplication and summation operations. Thus, the merged features can be obtained through a linear transformation.

\subsection{Pipeline of our Framework}\label{pipline}

In the pipeline of the considered DL network, eight consecutive images are inputted and utilized. As each is equivalent to a video clip, they contain motion information, which is helpful for the beam prediction. Combined with the visual information from each image, location, motion, and blockage information can be extracted from these RGB images. The pre-trained 3D ResNext with 101 layers (3D ResNext101) is adopted to extract motion features and the pre-trained ResNet with 152 layers (ResNet152) is used to extract visual features. These features are then merged as a vector through FFM and sent to the predictive network. 

As depicted in Fig. \ref{fig:lstms}, there are three kinds of inputs in the predictive network, namely, initial state, embedded beam vectors, and merged feature vector. The initial state is set as a vector of all zeros. Embedding is mapping a constant (beam index) to a vector and can represent the relation between constants well. So the embedded vector is utilized to represent the beam index. In each LSTM cell, the embedded beam vector and the merged feature vector are firstly transformed to the same shape and then sum them up as the `event'. According to the event, short and long term memories are obtained from the previous LSTM cell, and each cell predicts future 1 output vector whose index of the maximum element is the beam index. Notably, all the LSTM cells share the same merged features.

Based on the 1D and 2D LSTM networks introduced in Section \ref{lstms}, three methods are proposed and explained below.

\renewcommand{\arraystretch}{1.5}
  \begin{table*}[!ht]
 \centering
 \caption{Performance of Top-1 Accuracy and Exponential Decay Scores }
 \begin{adjustbox}{max width=\textwidth}
\begin{tabular}{c|c|c|c|c|c|c|c}
\hline\hline
& \multicolumn{3}{c|}{Top-1 Accuracy} & \multicolumn{3}{c|}{Exponential Decay Score} & \multirow{2}{*}{\makecell[c]{Running Time (s) \\ for `1 future beam' prediction}}\\ \cline{1-7}
& 1 future beam & 3 future beams & 5 future beams & 1 future beam & 3 future beams & 5 future beams \\ \hline
Method with 1D LSTM Network & 0.9170   & 0.7719   & 0.6448& 0.9238   & 0.8206   & 0.7356 &{0.42+0.016}     \\ \hline

Method with Modified 1D LSTM Network  & 0.8887   & 0.6260   & 0.4800 & 0.8974   & 0.7137   & 0.6129&{0.42+0.024}      \\ \hline
Method with 2D LSTM Network & 0.8704   & 0.5893   & 0.4503 & 0.8803   & 0.6857   & 0.5877  &{0.42+0.064 }   \\ \hline

Baseline Method in \cite{alrabeiah2020viwi}  & 0.85& 0.60& 0.50& 0.86& 0.68& 0.60&{0.0016}\\ \hline\hline

\end{tabular}\label{accuracyscore}
\end{adjustbox}
\end{table*}

\begin{figure}[!htbp]
    \centering
    \setlength{\abovecaptionskip}{0pt}
    \setlength{\belowcaptionskip}{10pt}
    \includegraphics[width=2.8 in]{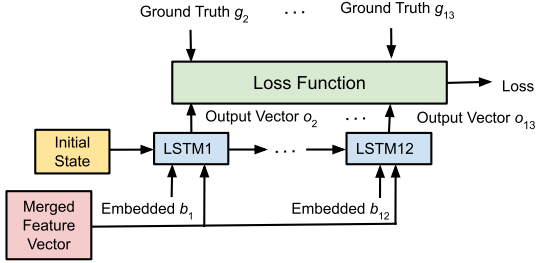}
    \caption{Training procedure of method with 1D LSTM netowrk}
    \label{fig:train1d}\vspace{-2mm}
\end{figure}
\subsubsection{Method with 1D LSTM Network}
When the predictive network in Fig. \ref{fig:framework} is a 1D LSTM network, the first method is obtained, as presented in Fig. \ref{fig:lstms}. The LSTM cell is recursive 12 times. The cell at the $k$th moment is denoted as the `$k$th LSTM cell'.

As shown in Fig. \ref{fig:train1d}, during the training process, the pipeline of our first method is the following:

\begin{enumerate}[itemindent=2.6em]
    \item[\textbf{Step 1:}] Eight consecutive images are fed to the pre-trained ResNet152 and 3D ResNext101 and then visual features and motion features are obtained;
    \item[\textbf{Step 2:}] These features from \textbf{step 1} are merged through the FFM;
    \item[\textbf{Step 3:}] The output vector from the FFM is fed to each LSTM cell as an input;
    \item[\textbf{Step 4:}] The embedded vectors of the first 12 beam indices go through the first to the last LSTM cells to update the hidden states and generate 12 output vectors;
    \item[\textbf{Step 5:}] The 12 output vectors are used to calculate the training loss with the ground truth and train the network.
\end{enumerate}

\begin{figure}[!htbp]
    \setlength{\abovecaptionskip}{0pt}
    \setlength{\belowcaptionskip}{10pt}
    \includegraphics[width=3.5 in]{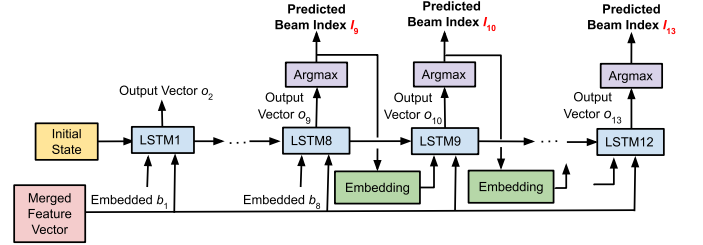}
    \caption{Testing procedure of method with 1D LSTM netowrk}
    \centering
    \label{fig:test1d}\vspace{-2mm}
\end{figure}
During the testing process, as we only have the first eight beam indices and images, the fourth step above is not applicable and is separated into two sub-steps as depicted in Fig. \ref{fig:test1d}:

\begin{enumerate}[itemindent=5em]
    \item[\textbf{Substep 4.1:}] The embedded vectors of the first eight beam indices go through the first to seventh LSTM cells and update the hidden states;
    \item[\textbf{Substep 4.2:}] The eighth to twelfth LSTM cells are used to predict the future beam indices which are obtained by acquiring the indices of the maximum element in these output vectors. Each cell is fed with the hidden state and the embedded beam index from the prediction of the previous LSTM.
\end{enumerate}

The fifth step is skipped during testing.

\subsubsection{Method with Modified 1D LSTM Network}
In our first method, the training and testing procedures are different. Actually, the first method essentially aims to predict the next beam index as we utilize all the first 12 beam indices as inputs during the training. During the testing process, among the eighth to twelfth predicting beam indices, the previous one's correct prediction is important for the next prediction. To make training and testing processes consistent, we designed a modified version of the first method, in which the output vector of each of the last five LSTM cells undergoes a linear transformation module and is fed to the next cell as the embedded input. In this way, only the first eight beam indices are used as inputs, and the training and testing can be the same.

\subsubsection{Method with 2D LSTM Network}
When we apply the 2D LSTM network to the predictive network, the third method can be realized as shown in Fig. \ref{fig:lstms}. In this method, we need to input the initial state, the merged feature vector and embedded vectors of the first eight beam indices into the LSTM network, and then get five outputs vectors directly. The training process is the same as the testing one. 

\subsection{Experiment}
In this section, we evaluate our three proposed methods on the ViWi-BT dataset. All the experiments are conducted in the framework of PyTorch on one NVIDIA V100 GPU. 
\subsubsection{Dataset}\label{dataset}
The VIWI-BT dataset contains a training set with 281,100 samples, a validation set with 120,468 samples, and a test set with 10,000 samples. There are 13 pairs of consecutive beam indices and corresponding images of street views in each sample of the training and validation sets. Furthermore, the first eight pairs are the observed beams for the target user and the sequence of the images where the target user appears. The last five pairs are groundtruth data containing the future beams and images of the same user. In this experiment, the first eight pairs serve as the inputs of the designed DL network to generate the predicted future five beam indices which are compared with the groundtruth ones.

\subsubsection{Implementation Details}

We first use the pre-trained ResNet152 and 3D ResNext101 to extract 2048-dimensional visual and 8192-dimensional motion features from the first eight images of each sample. The merged features are embedded as a 463-dimensional vector and fed to the predictive LSTM network. There are a 512-dimensional hidden size and a 129-dimensional output vector in each LSTM cell. The training pipeline mentioned in Section \ref{pipline} is then implemented to train the proposed network. 

 During the training, the designed DL network is optimized by the Adam optimizer. The learning rate is set as $4\times 10^{-4}$ at first and reduced by half every eight epochs. The batch size is set as 256. The cross-entropy loss is utilized for the loss function. 
 
 \subsubsection{Performance}

Following the evaluation in \cite{alrabeiah2020viwi}, the performances of our proposed methods are evaluated on the validation set with the same metrics, which are the top-1 accuracy and exponential decay score. 

{As defined in \cite{alrabeiah2020viwi}, the top-1 accuracy of $n$ future beams is expressed as
\begin{align}
    Acc^{(n)}_{\rm{top\text{-}1}}=\frac{1}{M}\sum\limits_{i=1}^{M}\mathbbm{1}\{\mathbf{\Hat{g}}^{(n)}_i=\mathbf{g}^{(n)}_i\},
\end{align}
where $M$ is the number of the samples in the validation set, $\mathbbm{1}\{\cdot\}$ denotes the indicator function, and $\mathbf{\Hat{g}}^{(n)}_i$ and $\mathbf{g}^{(n)}_i$ represent the predictive and groundtruth beam indices vectors of the $i$th sample with the length $n$.}

{The exponential decay score of $n$ future beams is given as
\begin{align}
    {\rm{score}}^{(n)}=\frac{1}{M}\sum\limits_{i=1}^{M}\exp{\left(-\frac{||\mathbf{\Hat{g}}^{(n)}_i-\mathbf{g}^{(n)}_i||_1}{n\sigma}\right)},
\end{align}
where $\sigma=0.5$ is a penalization parameter. }

Table \ref{accuracyscore} lists our results, in which the baseline method in \cite{alrabeiah2020viwi} is considered for comparison purposes. In the baseline method, the authors simply leveraged the beam-index data and ignored image data.

From the top-1 accuracy, we can see that our proposed method with the 1D LSTM network outperforms the baseline method in \cite{alrabeiah2020viwi}. The
method with a modified 1D LSTM network is better than the baseline method on '1 future beam' and '3 future beams'. The method with only the 2D LSTM network performs better than the baseline method on '1 future beam'.

For the exponential decay scores, the designed methods with the 1D LSTM network and modified 1D LSTM network absolutely outperformed the baseline method. The method with the 2D LSTM network is better than the baseline on `1 future beam' and `3 future beams' but a little worse on `5 future beams'.

{Our proposed methods outperform the baseline method on predicting `1 future beam'. Because the location, blockage, and speed information of the target user is extracted from the images and represented as motion and visual features to assist the prediction, and advanced LSTM networks are leveraged as the predictive networks. Among the three proposed methods, the method with the 1D LSTM network shows the best beam prediction performance in the target mobile scenarios. Compared with this method, the other two exhibit extra linear transformation modules or more LSTM cells in their predictive networks and need to be trained by more data. Therefore, performance degradation occurring on the predictions of `3 future beams' and `5 future beams' are caused by the small size of the training dataset. }

{The computational complexity here is measured by the running time of `1 future beam' prediction which exhibits the best performance and is more potential to be implemented in the practical wireless communication systems.} The running time of our methods consists of the execution time of feature extraction, FFM, and predictive network. It takes 0.42 seconds for the pre-trained 3D ResNext101 and ResNet152 to extract features from each set of eight images. Our method with 2D LSTM network exhibits the longest average running time due to its most complex structure shown in Fig. 5. The baseline method runs for the shortest time as it utilizes simple GRUs as the predictive network and abandons the image data. The method with 1D LSTM network shows the best predictive performance but a moderate prediction time, 0.016 seconds.

{The feature extraction takes a little long time which will cause latency issues but can be mitigated by employing more efficient CNNs in future work. From Table 1 of \cite{liu2020teinet}, we can see that the TSN \cite{wang2016temporal} takes 15.5 ms and the ECO \cite{zolfaghari2018eco} takes 17.4 ms on NVIDIA Tesla P100 GPU to extract 3D spatiotemporal features from eight images. In practical application, only one new image is captured at each time instance, and the previously extracted spatiotemporal features can be merged with the new image to reduce the latency. Similarly, for the 2D feature extraction, only the new image needs to be processed at each time instance. Table 2 of \cite{farhadi2018yolov3} shows that the Darknet-19 can achieve a speed of 171 frames per second (5.85 ms per image) to extract 2D features on NVIDIA Titan X GPU. 2D and 3D features extractions can be conducted simultaneously which means that the longer latter determines the whole feature extraction time. By jointly using TSN and Darknet-19, the running time of the method with 1D LSTM network can be reduced to less than 15.5+16=31.5 ms on V100 GPU which is more powerful than P100 and Titan X. Besides, in our experiments, we built the LSTM cell by ourselves. If we use the basic building block of LSTM in PyTorch, the comparable running time of the predictive network can be obtained with the baseline method. In engineering practice, even shorter latency can be achieved by utilizing GPU computing in the whole process and designing a more efficient framework (platform).}

{\subsubsection{Practical Application}In practical scenarios, RGB images and their corresponding beam indices can be obtained by cameras installed on the BS and the classic beamforming algorithm. After obtaining sufficient data for the training set, our proposed network will be pre-trained on these data and then run in the processors of the BS for beamforming. At the beginning of the serving time, the first eight beam indices can be estimated by the classic beamforming algorithm. Then the eight pairs of images and beam indices are sent to the processors for future beam predictions. Notably, after the first eight beams, the following beams will be predicted by using previously-obtained images and beam indices. These predicted beam indices and their corresponding images can be added to the training set to enlarge the dataset and improve the performance.}

\section{Challenges and Open Problems}
Although the previous sections elaborated on leveraging CV to tackle the mmWave beamforming problem, some challenges and open problems still exist in the front way of applying DL-based CV technologies in wireless communications.

\subsection{Building Datasets}

As DL is immensely data-hungry, a large dataset can guarantee the successful application of DL-based CV techniques to wireless communications. A qualified dataset in CV usually includes more than 10,000s of samples. For example, there are more than 14 million images in ImageNet, 60,000 images in Cifar-10, and roughly 650,000 video clips in Kinetics. It takes much time, money, and labor to generate such a huge amount of visual data. However, building a qualified dataset, which should be comprehensive and exhibit a balanced diversity of data, is still far from accomplished. These data should be able to represent all possibilities in the corresponding problem, and the amounts of different kinds of data can not make such a difference. Usually, a training set, a validation set, and a test set comprise a dataset. These three sets should be homogeneous and not overlapping, so randomly sampling them from a shuffled data pool is a better way to obtain the three sets. These data should be well-organized and easily manipulated. Therefore, the hardest work in DL is to build a satisfactory dataset.

\subsection{Selecting CV Techniques}
Many state-of-the-art DL techniques have been proved efficient and powerful in CV, such as reinforcement learning, encoder-decoder architecture, generative adversarial networks (GANs), Transformer, graph convolutional networks (GCNs), etc. Reinforcement learning has been widely applied in tackling optimization problems in wireless communications\cite{luong2019applications}. GCNs can be leveraged to address network-related issues\cite{rusek2019unveiling}, and encoder-decoder architecture is widely used in semantic segmentation and sequence-to-sequence tasks. The GAN is an immensely powerful CNN to learn the statistics of training data and has been widely used to improve the performance of other DL networks in CV\cite{goodfellow2016deep}. Transformer built on the attention mechanisms is a kind of the encoder-decoder architecture that can handle unordered sequences of data\cite{vaswani2017attention}. Much CV research has shown that if these techniques are jointly applied to make full use of the visual data, better results can be obtained\cite{wang2019controllable,park2019adversarial,hua2020gan}. 

Thus, a single proper CV technique or an adequate combination of several CV techniques are required to handle a specific problem in wireless systems. In the example given in Section \ref{sectionarchitecture}, we combined ResNet, 3D ResNext, and an LSTM network to achieve the required performance. Finding proper, efficient CV techniques thus remains an open problem. 

\subsection{Open Problems in Vision-Aided Wireless Communications}
\begin{figure}[!htbp]
    \centering
    \setlength{\abovecaptionskip}{0pt}
    \setlength{\belowcaptionskip}{10pt}
    \includegraphics[width=1.8 in]{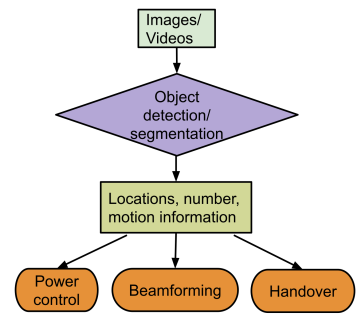}
    \caption{Pipeline of applying DL-based CV to cellular networks}
    \label{fig:cellular}\vspace{-2mm}
\end{figure}

\begin{figure}[!htbp]
    \centering
    \setlength{\abovecaptionskip}{0pt}
    \setlength{\belowcaptionskip}{10pt}
    \includegraphics[width=3.4 in]{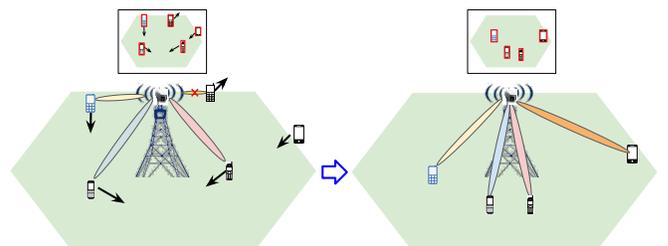}
    \caption{An example of applying DL-based CV to cellular networks}
    \label{fig:multiaccess}\vspace{-2mm}
\end{figure}
\begin{figure*}[!htbp]
    \centering
    \setlength{\abovecaptionskip}{0pt}
    \setlength{\belowcaptionskip}{10pt}
    \includegraphics[width=5 in]{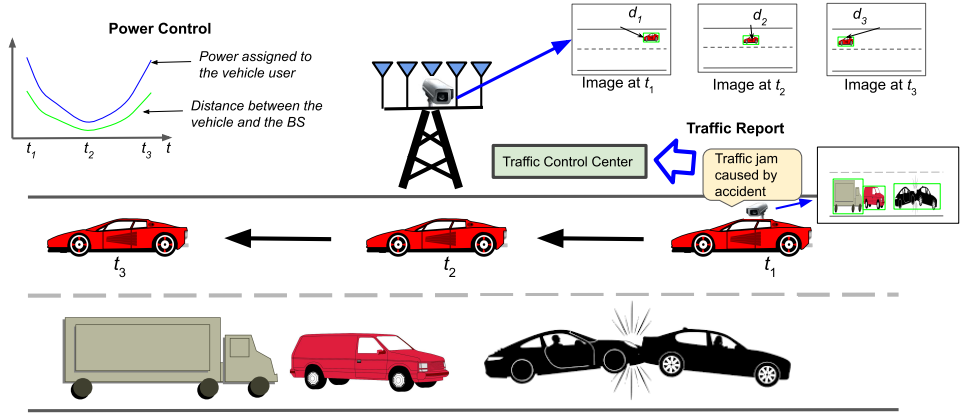}
    \caption{An example of applying DL-based CV to vehicle-to-everything communications }
    \label{fig:vehicle}\vspace{-2mm}
\end{figure*}

 As many kinds of cameras and LIDARs operate in real life, an enormous amount of visual data can be obtained through them, for more accurate motion and position information in the terminals that can be recognized, analyzed, and extracted from these multimedia data, which can also be explored to design and optimize wireless communications. Thus, some open problems in wireless communication scenarios are introduced and discussed as follows:

 (1) Cellular networks: As shown in Fig. \ref{fig:cellular}, visual data obtained at the BS in a cellular network may contain the locations, number, and motion information of the terminals in the open area. This information can be used for the BS to adjust its transmitting power and beam direction to save power consumption and reduce interference. Fig. \ref{fig:multiaccess} presents a real-life example: the motion information of the users in the coverage area of a BS can be utilized to forecast the future positions of these terminals and judge whether/when a terminal goes out or comes into its serving area. Then, transmit power and beam can be accurately assigned for these users who still stay in the coverage area, and channel resource allocation can be set up for the handover process to improve the utilization efficiency of the system resource.

 (2) Vehicle-to-everything communications: Visual data captured by one vehicle can reveal its environments, such as traffic conditions, which can be used to set up links with neighboring terminals, access points, and vehicles. Therefore, traffic schedules and jam/accident alarms can be conducted for improved road safety, traffic efficiency, and energy savings. As depicted in Fig \ref{fig:vehicle}, the BS located on the side of the road utilizes the visual information to estimate the distance of a vehicle terminal from it and adjust its transmit power accordingly for power saving and interference deduction purposes. Moreover, the images or videos captured by the cameras assembled on the vehicles can detect the traffic jam or accident, and then forwards the observed traffic information to the traffic control center for alarm and future traffic schedule.
 
  \begin{figure}[!htbp]
    \centering
    \setlength{\abovecaptionskip}{0pt}
    \setlength{\belowcaptionskip}{10pt}
    \includegraphics[width=2.8 in]{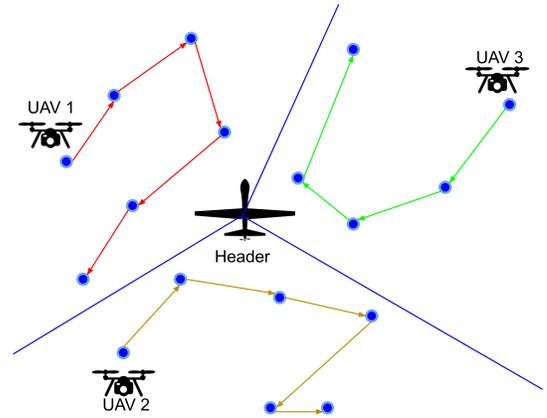}
    \caption{An example of applying DL-based CV to UAV-ground communications }
    \label{fig:uav}\vspace{-2mm}
\end{figure}
 (3) Unmanned aerial vehicle (UAV)-ground communications: When a UAV serves as an aerial BS, visual data captured by the UAV can be used to identify the locations and distribution of ground terminals, which can be utilized in power allocation, route/trajectory planning, etc. Moreover, when a ground BS communicates with several UAVs, visual data captured by the ground terminal can be used to define the serving range, allocate the channels/power, and so forth. Fig. \ref{fig:uav} illustrates an example in which a group of UAVs communicates with a set of ground terminals. The header UAV first takes an image of the whole area and detects all the terminals. Then the serving area is divided into several subareas. Each UAV serves one specific subarea and designs a route schedule according to the location information of these ground terminals obtained from the images. 
 
 (4) Smart cities: Visual data captured by satellites or airborne crafts can be applied to recognize and analyze the user's distribution and schedule power budget/serving ranges to achieve optimal energy efficiency.
   \begin{figure}[!htbp]
    \centering
    \setlength{\abovecaptionskip}{0pt}
    \setlength{\belowcaptionskip}{10pt}
    \includegraphics[width=3.4 in]{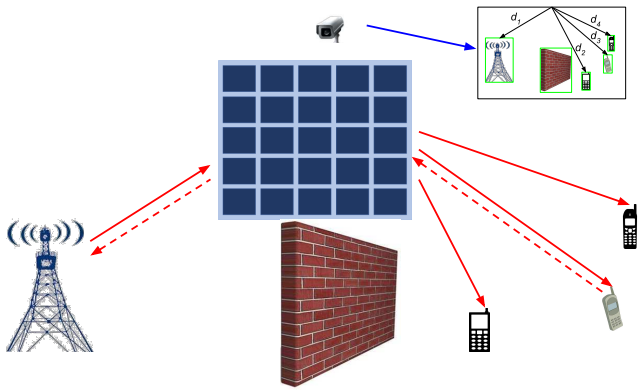}
    \caption{An example of applying DL-based CV to IRS system}
    \label{fig:irs}\vspace{-2mm}
\end{figure}

 (5) Intelligent reflecting surfaces (IRSs): Usually, implementing channel estimation and achieving network state information at an IRS is impossible because there is no comparable calculation capacity and no radio frequency (RF) signal transmitting or receiving capabilities at the IRS. Fortunately, DL-based CV can offer useful information to compensate for this gap. Thus, a proper control matrix can be optimally designed to accurately reflect the incident signals to the target destination by utilizing the visual data captured by the camera installed on the IRS, which includes the locations, distances, and the number of terminals shown in Fig. \ref{fig:irs}.

\section{Conclusion}

This article mainly presented the methodologies, opportunities, and challenges of applying DL-based CV to wireless communications. First, we discussed the feasibility of applying a DL-based CV in physical, MAC, and network layers in wireless communication systems. Second, we overviewed related datasets and work. Third, we gave an example of applying a DL-based CV to a mmWave MIMO beamforming system. In this example, previously observed images and beam indices were leveraged to predict future beam indices using ResNet, 3D ResNext, and an LSTM network. The experimental results show that visual data can significantly improve the accuracy of beam prediction. Finally, challenges and possible research directions were discussed. We hope this work stimulates future research innovations and fruitful results.

\bibliography{citation}


\end{document}